\documentclass[aps,prb,twocolumn,superscriptaddress,showpacs,showkeys,amsmath,amssymb]{revtex4}

\usepackage{epsfig,amsmath,amssymb,color}
\begin{document}

\title{Effective low-energy description of almost Ising-Heisenberg diamond chain}

\author{Oleg Derzhko}

\affiliation{Institute for Condensed Matter Physics,
National Academy of Sciences of Ukraine,
Svientsitskii Street 1, 79011 L'viv, Ukraine}

\affiliation{Department for Theoretical Physics,
Ivan Franko National University of L'viv,
Drahomanov Street 12, 79005 L'viv, Ukraine}

\affiliation{Abdus Salam International Centre for Theoretical Physics,
Strada Costiera 11, 34151 Trieste, Italy}

\author{Olesia Krupnitska}

\affiliation{Institute for Condensed Matter Physics,
National Academy of Sciences of Ukraine,
Svientsitskii Street 1, 79011 L'viv, Ukraine}

\author{Bohdan Lisnyi}

\affiliation{Institute for Condensed Matter Physics,
National Academy of Sciences of Ukraine,
Svientsitskii Street 1, 79011 L'viv, Ukraine}

\author{Jozef Stre\v{c}ka}

\affiliation{Department of Theoretical Physics and Astrophysics, Faculty of Science,
P.~J.~\v{S}af\'{a}rik University,
Park Angelinum 9, 040 01 Ko\v{s}ice, Slovak Republic}

\date{\today}

\begin{abstract}
We consider a geometrically frustrated spin-1/2 Ising-Heisenberg diamond chain, 
which is an exactly solvable model 
when assuming part of the exchange interactions as Heisenberg ones and another part as Ising ones. 
A small $XY$ part is afterwards perturbatively added to the Ising couplings, 
which enabled us to derive an effective Hamiltonian 
describing the low-energy behavior of the modified but full quantum version of the initial model. 
The effective model is much simpler and free of frustration. 
It is shown that the $XY$ part added to the originally Ising interaction gives rise to the spin-liquid phase 
with continuously varying magnetization, 
which emerges in between the magnetization plateaus and is totally absent in the initial hybrid diamond-chain model. 
The elaborated approach can also be applied to other hybrid Ising-Heisenberg spin systems.
\end{abstract}

\pacs{75.10.Jm; 75.10.Pa}

\keywords{Ising-Heisenberg spin systems, diamond chain}

\maketitle

Exactly solvable models, which can be solved without resorting to approximations,
are of great importance in statistical mechanics and condensed matter physics
since they provide milestones for our understanding of macroscopic properties of matter.\cite{baxter,mattis}
Well-known examples of such models are Ising models, vertex models, Bethe-ansatz models, etc.
Nowadays, numerous variations of Kitaev model\cite{kitaev} become quite popular.

Another way how to get exactly solvable models is to consider the so-called hybrid classical-quantum models,
which can be viewed as a regular pattern of small quantum parts linked through classical parts.
After making some transformations they can be mapped onto a simpler fully classical models 
with known exact solutions.\cite{syozi,fisher,strecka,jozef}

\begin{figure}
\begin{center}
\includegraphics[clip=on,width=80mm,angle=0]{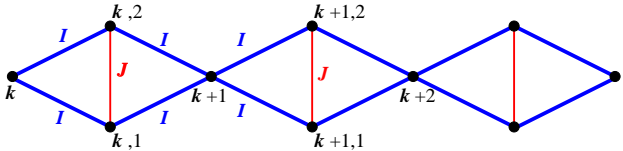}
\caption{(Color online)
A schematic representation of the frustrated Ising-Heisenberg diamond chain 
with the Ising coupling $I$ and the Heisenberg coupling $J$.
In the present study, a small $XY$ part $\propto\delta$, $\delta<1$ is added to the Ising coupling $I$ 
within the perturbative approach.}
\label{f01}
\end{center}
\end{figure}

One illustrating example of such a hybrid classical-quantum model is an Ising-Heisenberg diamond chain,
see Fig.~\ref{f01}. 
This model takes into account the Heisenberg interaction $J$ along the vertical bonds
and the Ising interaction $I$ along the bonds forming diamond motifs. 
In other words, the quantum Heisenberg spins 
($s$-spins) 
are situated at the bottom and top sites ($k,1$ and $k,2$) of the vertical bonds 
and the Ising spins 
($\mu$-spins) 
are placed at the connecting interstitial sites $k$, 
$k=1,\ldots,{\cal{N}}$, 
and $N=3{\cal{N}}$ is the total number of sites in the diamond chain, 
see Fig.~\ref{f01}. 
Over the last two decades, 
various versions of the spin-1/2 Heisenberg diamond chain have been extensively studied 
within different approaches.\cite{diamond,ijmpb}
The considerable interest aimed at this model has been stimulated 
by several solid-state realizations of the spin-1/2 Heisenberg diamond chain.\cite{kikuchi,fujihala} 
It is noteworthy that the simplified Ising-Heisenberg diamond-chain model 
allows a complete rigorous solution for all equilibrium properties.\cite{lucia1,lucia2,bohdan,lisnyi}
Although hybrid models can be also found among solid-state systems,\cite{ssr-of-hm} 
it seems more plausible to find in real life a case around the exactly solvable point. 
In the present study we will therefore address a question 
how to describe {\it{almost}} hybrid classical-quantum spin models, 
in which the Ising couplings acquire a small $XY$ part. 
The developed approach will be illustrated on a particular example of the frustrated spin-1/2 diamond-chain model, 
for which we will derive an effective Hamiltonian within the many-body perturbation theory. 
This description reproduces a low-temperature behavior of the initial model in a certain region of external magnetic fields.

To be more specific, 
we consider the frustrated spin-1/2 Ising-Heisenberg diamond chain 
(see Fig.~\ref{f01}) 
defined through the Hamiltonian
\begin{eqnarray}
\label{001}
H_0=\sum_{k=1}^{{\cal{N}}}H_{k,k+1},
\nonumber\\
H_{k,k+1}=
I\left(\mu_k^z+\mu_{k+1}^z\right)\left(s_{k,1}^z+s_{k,2}^z\right)
+J{\bf{s}}_{k,1}\cdot{\bf{s}}_{k,2}
\nonumber\\
-h\left(\mu_k^z+s_{k,1}^z+s_{k,2}^z\right).
\end{eqnarray}
This model is exactly solvable \cite{lucia1,lucia2,bohdan,lisnyi} 
and all thermodynamic quantities can be in principle calculated quite rigorously.
In particular, 
the ground state of the spin-1/2 Ising-Heisenberg diamond chain is known too 
and it basically depends on a relative ratio between the coupling constants $I$ and $J$. 
In what follows we will consider the special case $I>J>0$ for illustration.

Let us deviate from the exactly solvable point by adding a small $XY$ part to the originally Ising coupling, 
i.e.,
$H_0\to H= H_0 +V$
with
\begin{eqnarray}
\label{002}
V=\sum_{k=1}^{\cal{N}}V_{k,k+1},
\nonumber\\
V_{k,k+1}
=
\frac{\delta I}{2}
\left[
\left(\mu_{k}^+ + \mu_{k+1}^+\right)\left(s_{k,1}^- + s_{k,2}^-\right)
\right.
\nonumber\\
\left.
+\left(\mu_{k}^- + \mu_{k+1}^-\right)\left(s_{k,1}^+ + s_{k,2}^+\right) 
\right],
\end{eqnarray}
and $\delta<1$ is a small parameter.
Our aim is to construct an effective Hamiltonian for the low-energy properties of $H$ 
which is not too far from $H_0$.
The construction depends on the set of relevant ground states of $H_0$ 
which, in turn, depends on an external magnetic field $h$.

We begin with the case of {\it{high magnetic fields}}.
The ground state of $H_0$ above the saturation field reads
\begin{eqnarray}
\label{003}
\ldots
\left(\vert\uparrow\rangle\vert\uparrow_1\uparrow_2\rangle\right)_{k-1}
\left(\vert\uparrow\rangle\vert\uparrow_1\uparrow_2\rangle\right)_k
\left(\vert\uparrow\rangle\vert\uparrow_1\uparrow_2\rangle\right)_{k+1}
\ldots
\end{eqnarray}
with the energy $(I+J/4-3h/2){\cal{N}}$.
The ground state of $H_0$ below the saturation field reads
\begin{eqnarray}
\label{004}
\ldots
\left(\vert\downarrow\rangle\vert\uparrow_1\uparrow_2\rangle\right)_{k-1}
\left(\vert\downarrow\rangle\vert\uparrow_1\uparrow_2\rangle\right)_k
\left(\vert\downarrow\rangle\vert\uparrow_1\uparrow_2\rangle\right)_{k+1}
\ldots
\end{eqnarray}
with the energy $(-I+J/4-h/2){\cal{N}}$.
The saturation field $h_{{\rm{sat}}}$ follows from the equation
$I+J/4-3h_{{\rm{sat}}}/2=-I+J/4-h_{{\rm{sat}}}/2$
resulting in
$h_{{\rm{sat}}}=2I$.
At the saturation field the ground state is $2^{{\cal{N}}}$-fold degenerate:
Each $\mu$-spin may be directed either up $\uparrow$ or down $\downarrow$ 
without change of the ground-state energy $\varepsilon_0=(-2I+J/4){\cal{N}}$.

Let us introduce the projector on the ground-state manifold of $H_0$ precisely at $h=h_{{\rm{sat}}}$
\begin{eqnarray}
\label{005}
P=\vert\varphi_0\rangle\langle\varphi_0\vert
=
\otimes_{k=1}^{{\cal{N}}}
\left(\vert u\rangle\langle u\vert + \vert d\rangle\langle d\vert\right)_{k},
\nonumber\\
\vert u\rangle=\vert\uparrow\rangle\vert\uparrow_1\uparrow_2\rangle,
\;\;\;
\vert d\rangle=\vert\downarrow\rangle\vert\uparrow_1\uparrow_2\rangle.
\end{eqnarray}
It is also convenient to introduce the following (pseudo)spin-1/2 operators
\begin{eqnarray}
\label{006}
T^z_{k}=\frac{1}{2}\left(\vert u\rangle\langle u\vert - \vert d\rangle\langle d\vert\right)_{k},
\nonumber\\
T^+_{k} = \left(\vert u\rangle\langle d\vert\right)_{k},
\;\;\;
T^-_{k} = \left(\vert d\rangle\langle u\vert\right)_{k},
\nonumber\\
T^x_{k} = \frac{1}{2}\left(T^+_{k}+T^-_{k}\right),
\;\;\;
T^y_{k} = \frac{1}{2{\rm{i}}}\left(T^+_{k}-T^-_{k}\right).
\end{eqnarray}

It is quite evident that $2^{{\cal{N}}}$-fold degeneracy of $H_0$ is lifted 
under deviation from $h=h_{{\rm{sat}}}$ and switching on the perturbation $V$ given by (\ref{002}). 
There is an effective Hamiltonian ${\sf{H}}$ acting in this subspace, 
which yields the low-energy properties of the initial model $H_0+V$ in the high-field regime.
The effective Hamiltonian can be calculated perturbatively according to the formula
\begin{eqnarray}
\label{007}
{\sf{H}}
=
PHP
+PV\sum_{\alpha\ne 0}\frac{\vert\varphi_\alpha\rangle\langle\varphi_\alpha\vert}{\varepsilon_0-\varepsilon_\alpha}VP
+\ldots,
\end{eqnarray}
where $\vert\varphi_\alpha\rangle$, $\alpha\ne 0$ denotes the excited states of $H_0$ at $h=h_{{\rm{sat}}}$,
see, e.g., Ref.~\onlinecite{fulde}.

Calculating each terms in the r.h.s. of Eq.~(\ref{007}) and using Eq.~(\ref{006})
(see Appendix~\ref{a} for details) 
we arrive at the following effective Hamiltonian
\begin{eqnarray}
\label{016}
{\sf{H}}=
{\cal{N}}{\sf{C}}
\nonumber\\
+\sum_{k=1}^{{\cal{N}}}
\left[-{\sf{h}}T_k^z
+{\sf{J}}\left(T^x_kT^x_{k+1}+T^y_kT^y_{k+1}\right) + {\sf{J}}^z T^z_kT^z_{k+1}\right],
\nonumber\\
{\sf{C}}=\frac{J}{4}-h - \frac{3\delta^2 I}{8},
\nonumber\\
{\sf{h}}=h-2I-\frac{\delta^2 I}{2},
\;\;\;
{\sf{J}}=-\delta^2 I,
\;\;\;
{\sf{J}}^z= \frac{\delta^2 I}{2}.
\end{eqnarray}
This is nothing but the spin-1/2 $XXZ$ Heisenberg chain in a longitudinal ($z$-directed) magnetic field.
Compared to the initial model, 
the effective model contains only ${\cal{N}}=N/3$ sites and is completely free of frustration.

Bearing this in mind, 
one may conclude 
that the model defined by the Hamiltonian $H=H_0+V$ [see Eqs.~(\ref{001}) and (\ref{002})] 
with $I>J>0$ around $h\approx 2I$ for $\delta<1$ 
is represented by the effective model given by the Hamiltonian $\sf H$ [see Eq.~(\ref{016})].
As a consequence, 
there exists a mapping relationship between the free energy of the initial and effective models
\begin{eqnarray}
\label{017}
-\frac{1}{N}T\ln {\rm{Tr}} \exp\left(-\frac{H_0+V}{T}\right)
\nonumber\\
\to
-\frac{1}{3}\frac{1}{{\cal{N}}}T\ln {\rm{Tr}} \exp\left(-\frac{{\sf{H}}}{T}\right),
\end{eqnarray}
which is valid for $h\approx 2I$, $\delta<1$ and sufficiently low temperatures.
According to Eq.~(\ref{017}), 
one can use a broad knowledge about the spin-1/2 $XXZ$ Heisenberg chain in a longitudinal field
to understand the behavior of a more complicated initial diamond-chain model (\ref{001}), (\ref{002}).

\begin{figure}
\begin{center}
\includegraphics[clip=on,width=80mm,angle=0]{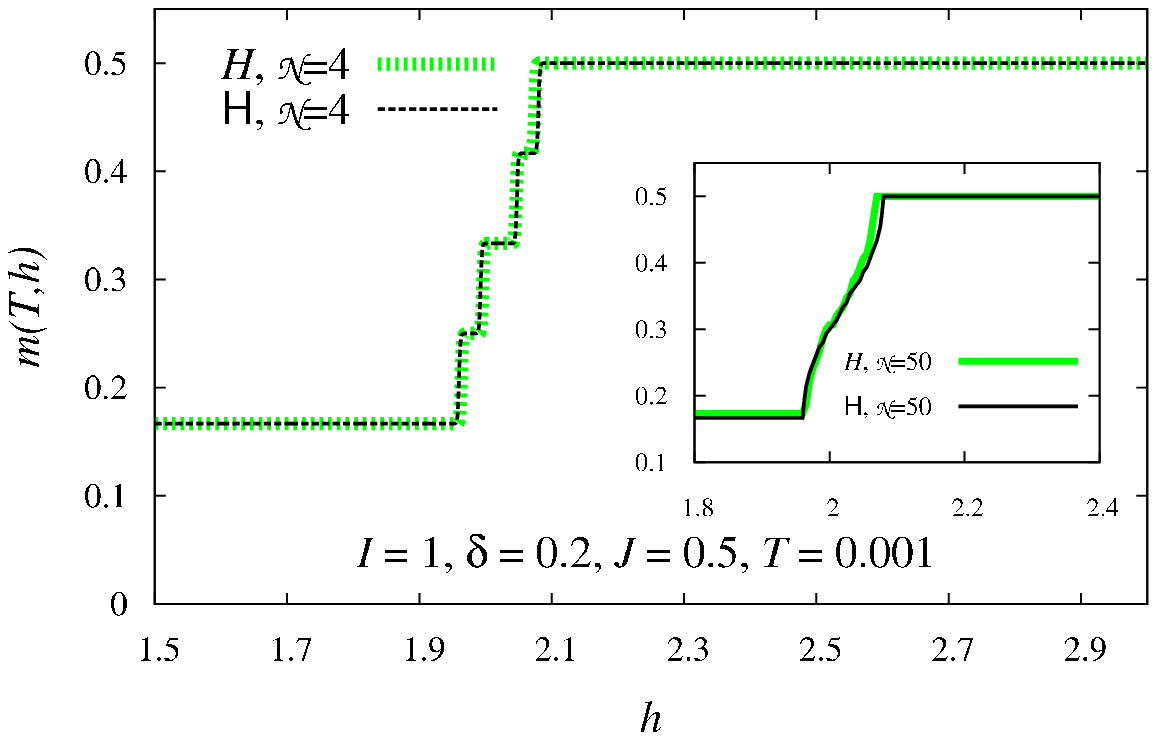}
\\
\vspace{2mm}
\includegraphics[clip=on,width=80mm,angle=0]{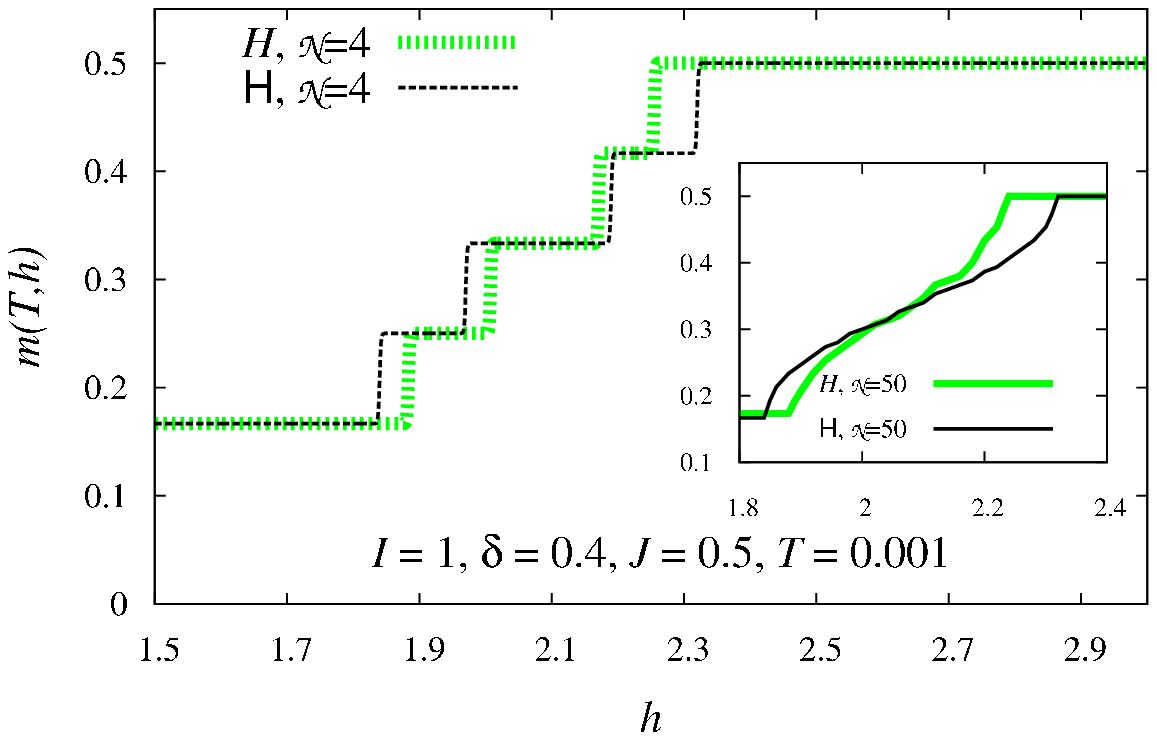}
\caption{(Color online)
High-field magnetization $m(T,h)$ vs $h$ at low temperatures $T$
for the initial model (\ref{001}), (\ref{002}) (bold green curves) 
and the effective model (\ref{016}) (thin black curves)
with $I=1$, $\delta=0.2$ (upper panel) and $\delta=0.4$ (lower panel), $J=0.5$, $T=0.001$.
ED data obtained for ${\cal{N}}=4$ (short-dashed curves),
DMRG data obtained for ${\cal{N}}=50$ (solid curves).}
\label{f02}
\end{center}
\end{figure}
\begin{figure}
\begin{center}
\includegraphics[clip=on,width=80mm,angle=0]{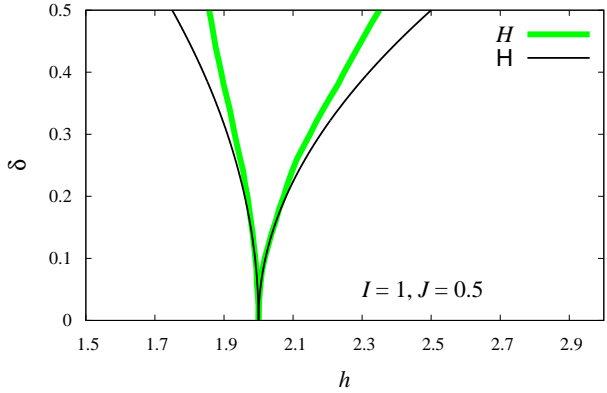}
\caption{(Color online)
Ground-state phase diagram in the plane $h$--$\delta$.
Spin-liquid phase appears in the region $h_l\le h\le h_h$ as $\delta$ deviates from zero
instead of the jump from 1/3-plateau state at $h=2I$ as $\delta=0$.
Effective model predictions 
(thin black solid curves)
are: 
$\delta(h)=\sqrt{2-h/I}$, $h\le 2I$ 
(left boundary for the spin-liquid phase)
and
$\delta(h)=\sqrt{2-h/I}/\sqrt{2}$, $h\ge 2I$ 
(right boundary for the spin-liquid phase).
Results for the initial model 
(bold green solid curves)
are obtained by DMRG calculations for open chains with ${\cal{N}}=50$.}
\label{f03}
\end{center}
\end{figure}
\begin{figure}
\begin{center}
\includegraphics[clip=on,width=80mm,angle=0]{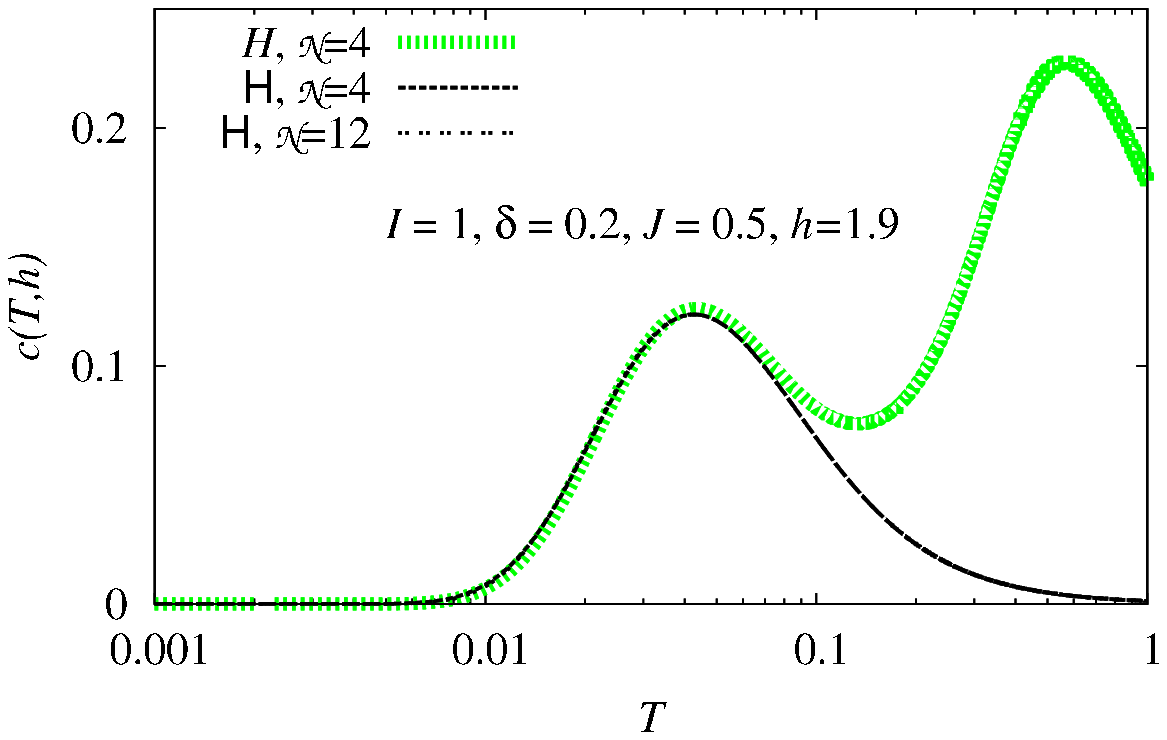}
\\
\vspace{2mm}
\includegraphics[clip=on,width=80mm,angle=0]{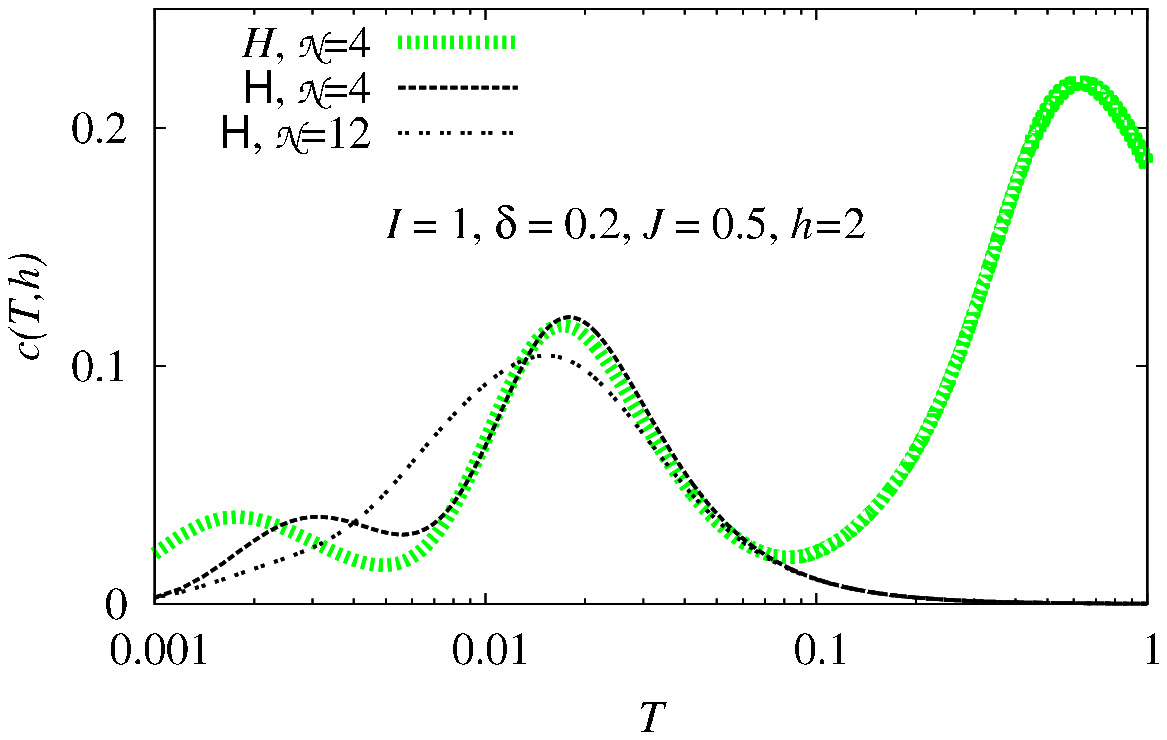}
\\
\vspace{2mm}
\includegraphics[clip=on,width=80mm,angle=0]{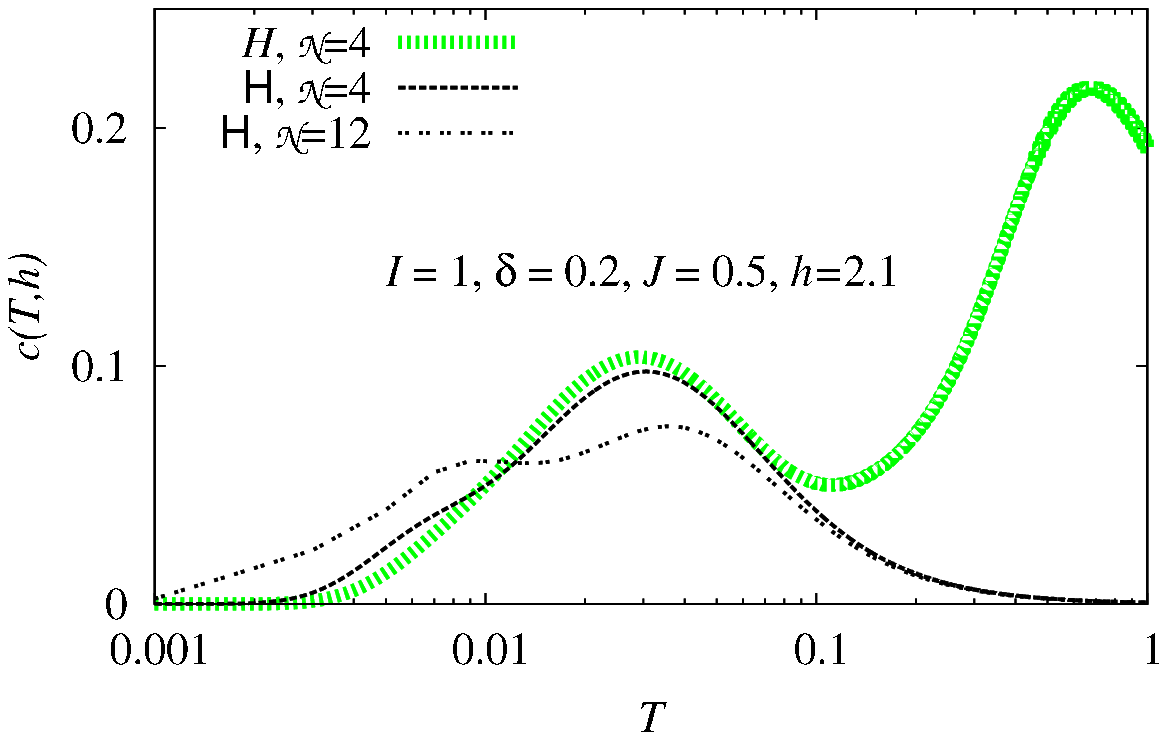}
\\
\vspace{2mm}
\includegraphics[clip=on,width=80mm,angle=0]{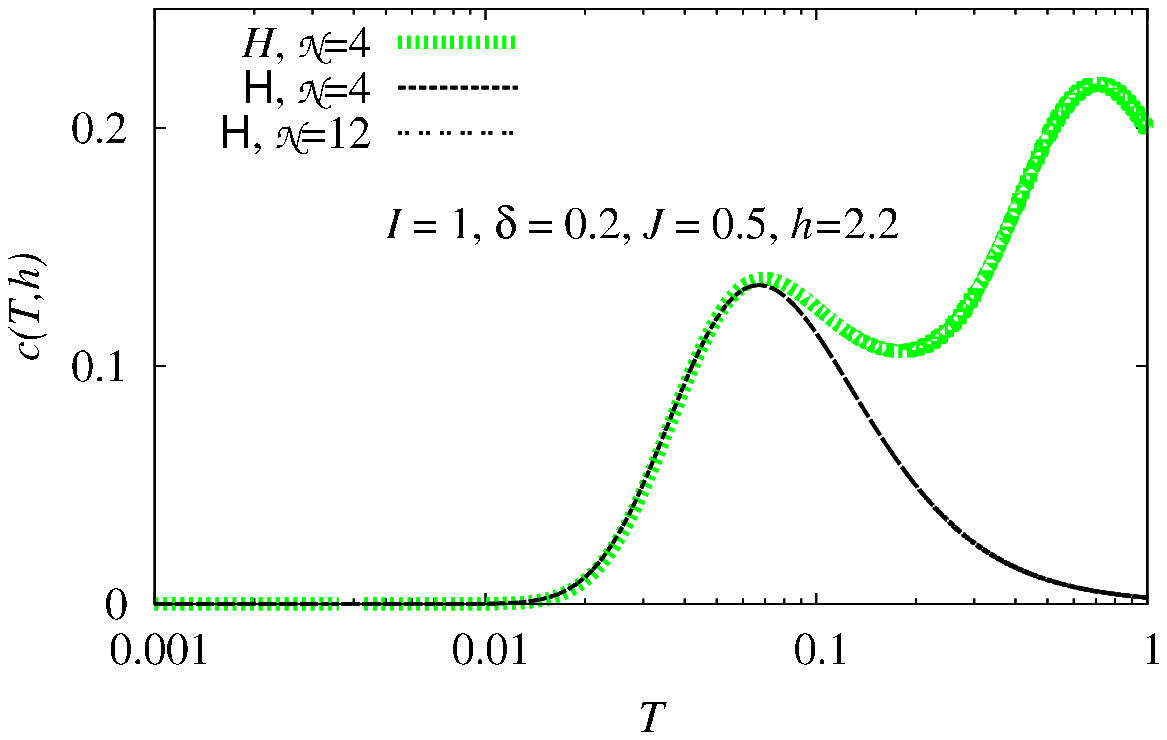}
\caption{(Color online)
Specific heat $c(T,h)$ vs $T$ at high fields $h$
for the initial model (\ref{001}), (\ref{002}) (bold green curves) 
and the effective model (\ref{016}) (thin black curves)
with $I=1$, $\delta=0.2$, $J=0.5$, $h=1.9,\,2,\,2.1,\,2.2$ (from top to bottom).
ED data reported for ${\cal{N}}=4$ (short-dashed curves) and ${\cal{N}}=12$ (double-dashed curves).}
\label{f04}
\end{center}
\end{figure}

This conclusion can be checked numerically 
performing exact-diagonalization (ED) and density-matrix-renormalization-group (DMRG) calculations.\cite{alps}
We consider 
periodic chains with ${\cal{N}}=4,\ldots, 12$ in ED calculations 
or 
open chains with ${\cal{N}}=50$ in DMRG calculations,
set $I=1$, $\delta=0.2\ldots 0.5$, $J=0.5$, $h=1.5\ldots 3$ 
and compute 
the magnetization per site of the initial chain $m(T,h)$ at low temperatures 
and 
the specific heat per site of the initial chain $c(T,h)$ at high fields for both models, 
see Figs.~\ref{f02}, \ref{f03}, and \ref{f04}.

As one can see from Fig.~\ref{f02}, 
the magnetization jump observable for $\delta=0$ in a zero-temperature magnetization curve at $h=2I$ 
between the one-third of the saturated magnetization $m_{{\rm{sat}}}=m(0,\infty)=1/2$ 
and the saturated magnetization $m_{{\rm{sat}}}$ 
transforms into a region $h_{l}\le h\le h_{h}$ with continuously varying magnetization 
whenever $\delta$ deviates from zero. 
The effective model reproduces this behavior reasonably well up to $\delta \approx 0.3$ 
and it gives evidence for a presence of the gapless spin-liquid phase before the magnetization reaches saturation. 
It is quite clear from Fig.~\ref{f02} 
that the effective theory slightly overestimates a magnetic-field range 
$h_{l}=2I-\delta^2 I$ and $h_{h}=2I +2\delta^2 I$ 
pertinent to continuous change of the magnetization within the spin-liquid regime. 
The finite-size effects present in ED data 
(main panels in Fig.~\ref{f02}) 
can be simply estimated from a comparison with DMRG data 
(insets in Fig.~\ref{f02}). 
In Fig.~\ref{f03} we present the ground-state phase diagram in the $h$--$\delta$ plane, 
which was obtained from DMRG simulations for the initial model and the analytical calculations for the effective model. 
Apparently, 
the magnetic-field range corresponding to a spin-liquid regime generally broadens upon increasing of the parameter $\delta$, 
which means that the $XY$-part of the originally Ising coupling is a primary cause for an existence of the spin-liquid phase.

Temperature dependences of the specific heat at high enough fields are shown in Fig.~\ref{f04} 
in order to demonstrate up to which temperature the elaborated effective description remains valid. 
For the selected set of parameters it holds until the temperature does not exceed approximately 0.08 
(in units $I=1$).
Note that a low-temperature discrepancy between the initial model and the effective model seen at $h=2.0$ and $h=2.1$ 
is a finite-size artifact, 
because finite-size effects are strong if the energy spectrum is gapless 
in contrast to the cases with gapped energy spectrum, 
cf. the panels for $h=2.0,\,2.1$ and for $h=1.9,\,2.2$ in Fig.~\ref{f04}.

In the case of {\it{low magnetic fields}}, 
the ground state of $H_0$ for infinitesimally small $h>0$ is
\begin{eqnarray}
\label{018}
\ldots
\left(\vert\downarrow\rangle\vert\uparrow_1\uparrow_2\rangle\right)_{k-1}
\left(\vert\downarrow\rangle\vert\uparrow_1\uparrow_2\rangle\right)_k
\left(\vert\downarrow\rangle\vert\uparrow_1\uparrow_2\rangle\right)_{k+1}
\ldots
\end{eqnarray}
with the energy $(-I+J/4-h/2){\cal{N}}$. 
The ground state of $H_0$ for infinitesimally small $h<0$ is
\begin{eqnarray}
\label{019}
\ldots
\left(\vert\uparrow\rangle\vert\downarrow_1\downarrow_2\rangle\right)_{k-1}
\left(\vert\uparrow\rangle\vert\downarrow_1\downarrow_2\rangle\right)_k
\left(\vert\uparrow\rangle\vert\downarrow_1\downarrow_2\rangle\right)_{k+1}
\ldots
\end{eqnarray}
with the energy $(-I+J/4+h/2){\cal{N}}$. 
At $h=0$ the ground state is 2-fold degenerate 
and the effective Hamiltonian acting in this subspace is simply
\begin{eqnarray}
\label{020}
{\sf{H}}={\cal{N}}\left(-I+\frac{J}{4}-\frac{\delta^2 I}{2}-h T^z\right),
\end{eqnarray}
see Appendix~\ref{b}.

To summarize, 
we have perturbatively added a small $XY$ part to the pure Ising coupling 
when starting from the exactly solvable Ising-Heisenberg model 
with the goal to construct an effective low-energy description of the full quantum Heisenberg analog 
of this hybrid classical-quantum model. 
We have exemplified this idea 
by investigating the symmetric spin-1/2 Ising-Heisenberg diamond chain 
with all equal Ising interactions along the diamond sides. 
It is worthwhile to remark that the developed approach can be adapted for other hybrid classical-quantum models 
such as for instance the Ising-Heisenberg orthogonal-dimer chain 
or the Ising-Heisenberg model on Shastry-Sutherland lattice.\cite{taras}

The authors thank T.~Verkholyak for fruitful discussions.
O.~D. acknowledges financial support of the Abdus Salam International Centre for Theoretical Physics 
(Trieste, August, 2015).
He also acknowledges support of the Pavol Jozef \v{S}af\'{a}rik University in Ko\v{s}ice 
during the 13$^{{\rm{th}}}$ Czech and Slovak Conference on Magnetism CSMAG'07 
(Ko\v{s}ice, July 9 -- 12, 2007)
when an idea of the present study arose.
O.~K. thanks National Academy of Sciences of Ukraine for the Scholarship for Young Researches.
J.~S. acknowledges financial support by Slovak Research and Development Agency 
provided under contract Nos. APVV-0132-11 and APVV-14-0073.

\appendix
\section{Derivation of effective Hamiltonian (\ref{016})}
\label{a}

Here we provide details of the derivation of the effective Hamiltonian, Eq.~(\ref{016}).
Let us calculate each terms in the r.h.s. of Eq.~(\ref{007}).

For $H_0$ (\ref{001}) at $h=h_{{\rm{sat}}}=2I$ 
we have 
$H_0\vert\varphi_0\rangle=\varepsilon_0\vert\varphi_0\rangle$
resulting in
$PH_0P=\varepsilon_0$,
$\varepsilon_0={\cal{N}}(-2I+J/4)$.
Furthermore,
\begin{eqnarray}
\label{008}
-\left(h-h_{{\rm{sat}}}\right)\left(\mu_k^z+s_{k,1}^z+s_{k,2}^z\right)\vert u\rangle_k
\nonumber\\
=
-\frac{3}{2}\left(h-h_{{\rm{sat}}}\right)\vert u\rangle_k,
\nonumber\\
-\left(h-h_{{\rm{sat}}}\right)\left(\mu_k^z+s_{k,1}^z+s_{k,2}^z\right)\vert d\rangle_k
\nonumber\\
=
-\frac{1}{2}\left(h-h_{{\rm{sat}}}\right)\vert d\rangle_k,
\end{eqnarray}
and therefore we get 
\begin{eqnarray}
\label{009}
PH_0P
=\sum_{k=1}^{\cal{N}}
\left[
-2I+\frac{J}{4}
\right.
\nonumber\\
\left.
-\frac{3}{2}\left(h-2I\right)\left(\vert u\rangle\langle u\vert\right)_k
-\frac{1}{2}\left(h-2I\right)\left(\vert d\rangle\langle d\vert\right)_k
\right]
\end{eqnarray}
for arbitrary $h$.
Using Eq.~(\ref{006}), 
we write Eq.~(\ref{009}) as follows:
\begin{eqnarray}
\label{010}
PH_0P
=\sum_{k=1}^{\cal{N}}
\left[\frac{J}{4}-h-\left(h-2I\right)T^z_k\right].
\end{eqnarray}

Furthermore, 
by direct calculations we find
\begin{eqnarray}
\label{011}
V_{k,k+1}\vert u\rangle_k\vert u\rangle_{k+1}
= 0,
\nonumber\\
V_{k,k+1}\vert u\rangle_k\vert d\rangle_{k+1}
= \frac{\delta I}{\sqrt{2}}
\left(\vert\uparrow\rangle\vert t,0\rangle\right)_k\vert u\rangle_{k+1},
\nonumber\\
V_{k,k+1}\vert d\rangle_k\vert u\rangle_{k+1}
= \frac{\delta I}{\sqrt{2}}
\left(\vert\uparrow\rangle\vert t,0\rangle\right)_k\vert u\rangle_{k+1},
\nonumber\\
V_{k,k+1}\vert d\rangle_k\vert d\rangle_{k+1}
= \frac{\delta I}{\sqrt{2}}
\left[
\left(\vert\uparrow\rangle\vert t,0\rangle\right)_k\vert d\rangle_{k+1}
\right.
\nonumber\\
\left.
+\left(\vert\downarrow\rangle\vert t,0\rangle\right)_k\vert u\rangle_{k+1}
\right],
\end{eqnarray}
where 
$\vert t,0\rangle=(\vert\uparrow_1\downarrow_2\rangle + \vert\downarrow_1\uparrow_2\rangle)/\sqrt{2}$.
Importantly,
the energy of the state 
$\left(\vert\uparrow\rangle\vert t,0\rangle\right)_k\vert u\rangle_{k+1}$ 
is higher than the energy of the states
$\vert u\rangle_k\vert d\rangle_{k+1}$ and $\vert d\rangle_k\vert u\rangle_{k+1}$
by $I$,
whereas 
the energy of the states 
$\left(\vert\uparrow\rangle\vert t,0\rangle\right)_k\vert d\rangle_{k+1}$
and
$\left(\vert\downarrow\rangle\vert t,0\rangle\right)_k\vert u\rangle_{k+1}$
is higher than the energy of the state
$\vert d\rangle_k\vert d\rangle_{k+1}$
by $2I$;
this can be easily checked by direct calculations.

As is seen from Eq.~(\ref{011}),
a nontrivial result of acting by $V_{k,k+1}$ gives a new state of the $k$th cell.
From Eq.~(\ref{011}) one immediately concludes that
\begin{eqnarray}
\label{012}
PVP=0.
\end{eqnarray}
However, the second term in Eq.~(\ref{007}) may be nonzero.
Really,
\begin{eqnarray}
\label{013}
PV\sum_{\alpha\ne 0}\frac{\vert\varphi_\alpha\rangle\langle\varphi_\alpha\vert}{\varepsilon_0-\varepsilon_\alpha}VP
=
\sum_{k=1}^{\cal{N}}
\nonumber\\
\left\{
\frac{\left(\frac{\delta I}{\sqrt{2}}\right)^2}{-I}
\left[
\left(\vert u\rangle\langle u\vert\right)_k \left(\vert d\rangle\langle d\vert\right)_{k+1}
+
\left(\vert u\rangle\langle d\vert\right)_k \left(\vert d\rangle\langle u\vert\right)_{k+1}
\right.
\right.
\nonumber\\
\left.
\left.
+\left(\vert d\rangle\langle u\vert\right)_k \left(\vert u\rangle\langle d\vert\right)_{k+1}
+\left(\vert d\rangle\langle d\vert\right)_k \left(\vert u\rangle\langle u\vert\right)_{k+1}\right]
\right.
\nonumber\\
\left.
+\frac{\left(\frac{\delta I}{\sqrt{2}}\right)^2}{-2I}
2\left(\vert d\rangle\langle d\vert\right)_k \left(\vert d\rangle\langle d\vert\right)_{k+1}
\right\}.
\end{eqnarray}
After some manipulations and utilizing Eq.~(\ref{006}), 
Eq.~(\ref{013}) can be cast into
\begin{eqnarray}
\label{014}
PV\sum_{\alpha\ne 0}\frac{\vert\varphi_\alpha\rangle\langle\varphi_\alpha\vert}{\varepsilon_0-\varepsilon_\alpha}VP=
-\frac{\delta^2 I}{2}
\nonumber\\
\times
\sum_{k=1}^{\cal{N}}
\left[
\frac{3}{4} - T^z_k + 2\left(T^x_kT^x_{k+1}+T^y_kT^y_{k+1}\right) - T^z_kT^z_{k+1}
\right].
\end{eqnarray}

Combining Eqs.~(\ref{010}), (\ref{012}), and (\ref{014}) we obtain the effective Hamiltonian 
\begin{eqnarray}
\label{015}
{\sf{H}}=
\sum_{k=1}^{{\cal{N}}}
\left[\frac{J}{4}-h - \frac{3\delta^2 I}{8}
-\left(h-2I-\frac{\delta^2 I}{2}\right)T_k^z
\right.
\nonumber\\
\left.
-\delta^2 I\left(T^x_kT^x_{k+1}+T^y_kT^y_{k+1}\right) + \frac{\delta^2 I}{2}T^z_kT^z_{k+1}
\right],
\end{eqnarray}
i.e., 
${\sf{H}}$ as it is given in Eq.~(\ref{016}).

\section{Effective Hamiltonian around $h=0$}
\label{b}

In this appendix we derive effective Hamiltonian (\ref{020}).
We denote 
$\vert u\rangle
=
\ldots
\left(\vert\downarrow\rangle\vert\uparrow_1\uparrow_2\rangle\right)_k
\ldots$
and
$\vert d\rangle
=
\ldots
\left(\vert\uparrow\rangle\vert\downarrow_1\downarrow_2\rangle\right)_k
\ldots$
and introduce 
the projector
$P=\vert u\rangle\langle u\vert + \vert d\rangle\langle d\vert$
and
the (pseudo)spin-1/2 operators
$T^z=(\vert u\rangle\langle u\vert - \vert d\rangle\langle d\vert)/2$, 
$T^+=\vert u\rangle\langle d\vert$,
$T^-=\vert d\rangle\langle u\vert$,
cf. Eqs.~(\ref{005}) and (\ref{006}).
Repeating the arguments which lead to Eq.~(\ref{010}),
we get 
$PH_0P=\sum_{k=1}^{{\cal{N}}}(-I+J/4-h T^z)$.
Furthermore,
\begin{eqnarray}
\label{021}
V_{k,k+1}\vert u\rangle
= 
\frac{\delta I}{\sqrt{2}}
\left[
\ldots\left(\vert\uparrow\rangle\vert t,0\rangle\right)_k\ldots
\right.
\nonumber\\
\left.
+
\ldots\left(\vert\downarrow\rangle\vert t,0\rangle\right)_k\left(\vert\uparrow\rangle\vert \uparrow_1\uparrow_2\rangle\right)_{k+1}\ldots
\right],
\nonumber\\
V_{k,k+1}\vert d\rangle
=
\frac{\delta I}{\sqrt{2}}
\left[
\ldots\left(\vert\downarrow\rangle\vert t,0\rangle\right)_k\ldots
\right.
\nonumber\\
\left.
+
\ldots\left(\vert\uparrow\rangle\vert t,0\rangle\right)_k\left(\vert\downarrow\rangle\vert \downarrow_1\downarrow_2\rangle\right)_{k+1}\ldots
\right];
\end{eqnarray}
the energy of the excited states which appear in the r.h.s. of Eq.~(\ref{021}) is higher than the ground-state energy by $2I$.
Obviously,
$PVP=0$ 
and
\begin{eqnarray}
\label{022}
PV\sum_{\alpha\ne 0}\frac{\vert\varphi_\alpha\rangle\langle\varphi_\alpha\vert}{\varepsilon_0-\varepsilon_\alpha}VP
\nonumber\\
=
\sum_{k=1}^{\cal{N}}
\left[
\frac{\left(\frac{\delta I}{\sqrt{2}}\right)^2}{-2I}
2\vert u\rangle\langle u\vert
+
\frac{\left(\frac{\delta I}{\sqrt{2}}\right)^2}{-2I}
2\vert d\rangle\langle d\vert
\right]
\nonumber\\
=
\sum_{k=1}^{\cal{N}}
\left(-\frac{\delta^2 I}{2}\right).
\end{eqnarray}
Combining all together, we immediately get Eq.~(\ref{020}).

\end{document}